\documentclass[proceedings]{JHEP}
\usepackage{epsfig}  
\def\pl#1#2#3{{\it Phys. Lett.}
{\bf #1~B}~(#2)~#3}
\def\apjl#1#2#3{{\it Astrophys. J. Lett.}
{\bf #1}~(#2)~#3}
\def\apj#1#2#3{{\it Astrophys. Journal}
{\bf #1}~(#2)~#3}

\def\pr#1#2#3{{\it Phys. Reports}
{\bf #1}~(#2)~#3}
\def\mnras#1#2#3{{\it Mon. Not. Roy. Astr. Soc.}
{\bf #1}~(#2)~#3}
\def\n#1#2#3{{\it Nature}
{\bf #1}~(#2)~#3}

\def\anj#1#2#3{{\it Astron. J.}
{\bf #1}~(#2)~#3}
\def\grg#1#2#3{{\it Gen. Rel. Grav.}
{\bf #1}~(#2)~#3}

\conference{Trieste Meeting of the TMR Network on Physics 
beyond the SM}

\title{Degenerate or Hierarchical Neutrinos in 
Supersymmetric Inflation}

\author{G. Lazarides\\
Physics Division, School of Technology,\\
Aristotle University of Thessaloniki,\\ 
Thessaloniki 540 06, Greece\\
E-mail: \email{lazaride@eng.auth.gr}}

\abstract
{Two moderate extensions of the minimal supersymmetric 
standard model are considered. The first one includes 
a $U(1)_{B-L}$ gauge group, while the second is based 
on a left-right symmetric gauge group. In these models, 
hybrid inflation is `naturally' realized and the $\mu$ 
problem is solved via a Peccei-Quinn symmetry. 
Baryon number conservation is an automatic consequence 
of a R-symmetry. The baryon 
asymmetry of the universe is generated through a 
primordial leptogenesis. In the `$B-L$' case, 
neutrinos are assumed to acquire degenerate masses 
$\approx 1.5~{\rm{eV}}$ by coupling to $SU(2)_L$ 
triplet superfields, thereby providing the hot dark 
matter of the universe. In the `left-right' model, 
light neutrinos acquire hierarchical masses by the 
seesaw mechanism. They are taken from the small angle 
MSW resolution of the solar neutrino puzzle and the 
SuperKamiokande data. Maximal $\nu_{\mu}-\nu_{\tau}$ 
mixing, implied by the same data, is easily accommodated. 
The gravitino and baryogenesis constraints can be satisfied, 
in both models, with more or less `natural' values of the 
relevant coupling constants.}

\begin{document} 

\par
Despite its compelling properties, the minimal 
supersymmetric standard model (MSSM) leaves a number of 
fundamental physical issues unanswered. This clearly 
indicates that it must be part of a more basic theory. 
Some of the shortcomings of MSSM, which are relevant for 
the discussion here, are in order:
\begin{list}
\setlength{\rightmargin=0cm}{\leftmargin=0cm}
\item[{\bf i)}] Inflation cannot be implemented. 
\item[{\bf ii)}] There is no understanding of how the 
$\mu$ term, with 
$\mu \sim 10^{2}-10^{3}~$ GeV, arises.
\item[{\bf iii)}] Neutrinos remain massless and, thus, 
there are no neutrino oscillations in contrast to recent 
experimental evidence \cite{superk}.  
\item[{\bf iv)}] Although the lightest supersymmetric 
particle (LSP) of MSSM is a promising candidate for cold 
dark matter, hot dark matter cannot be accommodated with 
purely MSSM fields. It has become increasingly clear, 
however, that a combination of both cold and hot dark 
matter is required \cite{structure} to fit the data 
on large scale structure formation in the universe, 
especially in the case of zero cosmological constant 
($\Lambda=0$). 
\item[{\bf v)}] The observed baryon asymmetry of the 
universe (BAU) cannot be generated easily in MSSM 
(through the nonperturbative electroweak sphaleron 
processes).
\end{list}

\par
All these problems can be simultaneously resolved 
in moderate extensions of MSSM. Two such extensions 
are based on the gauge groups:
\begin{list}
\setlength{\rightmargin=0cm}{\leftmargin=0cm}
\item[{\bf ($a$)}] $G_{S}\times U(1)_{B-L}\equiv 
G_{B-L}$ ($G_{S}$ being the standard model gauge group) 
\cite{deg}.
\item[{\bf ($b$)}] $SU(3)_{c}\times SU(2)_{L}
\times SU(2)_{R}\times U(1)_{B-L}\equiv G_{LR}$ 
\cite{lss,hier}.
\end{list}

\par
In the $G_{B-L}$ case, light neutrino masses can be 
generated by including \cite{triplet,rsym} $SU(2)_{L}$ 
triplet pairs of superfields $T_a$, $\bar{T}_a$ 
($a=1,2,...,n$). It is then not inconceivable that these 
masses are degenerate and we will assume them to be so. 
The hot dark matter of the universe can, in this case, 
consist of light neutrinos without any incompatibility with 
atmospheric \cite{superk} and solar neutrino oscillations 
even with three neutrino species.

\par
In the case of the left-right symmetric gauge group 
$G_{LR}$, right handed neutrino superfields, $\nu^{c}$,
are present forming $SU(2)_{R}$ doublets with the 
$SU(2)_{L}$ singlet charged antileptons $E^{c}$.
Light neutrino masses are then generated via the well-known
seesaw mechanism and cannot be `naturally' degenerate. We, 
thus, take hierarchical light neutrino masses in this case 
which, being unable to provide the hot dark matter, are 
more appropriate for a universe with nonzero cosmological 
constant ($\Lambda\neq 0$) favored by recent observations 
\cite{lambda}. In fact, it has been shown \cite{lambdafit} 
that, for $\Lambda\neq 0$, cold dark matter alone can lead 
to a `good' fit of the cosmic background radiation and both 
the large scale structure and age of the universe data. 
Moreover, the possibility of improving this fit by adding 
light neutrinos as hot dark matter appears \cite{primack} 
to be rather limited. Note that neutrino masses could be 
hierarchical even for $\Lambda= 0$ provided that hot 
dark matter consists of some other particles (say axinos).

\par
The spontaneous breaking of $G_{B-L}$ to $G_{S}$, at a 
superheavy mass scale $M\sim 10^{16}~{\rm{GeV}}$, is 
achieved via the renormalizable superpotential 
\begin{equation}
W=\kappa S(\phi\bar{\phi}-M^2)~,  
\label{W}
\end{equation}
where $\phi,~\bar{\phi}$ is a conjugate pair of standard 
model singlet left handed superfields with $B-L$ charges 
equal to 1, -1 respectively, and $S$ is a gauge singlet left 
handed superfield. The coupling constant $\kappa$ and the 
mass parameter $M$ can be made positive by suitable phase 
redefinitions. In the $G_{LR}$ case, $\phi\bar{\phi}$
in Eq.(\ref{W}) is replaced by $l^c\bar l^{c}$, where 
$l^c$, $\bar l^{c}$ is a conjugate pair of $SU(2)_R$ 
doublet left handed superfields with $B-L$ charges equal 
to 1, -1 respectively ($\phi,~\bar{\phi}$ correspond 
to the neutral components of $l^c$, $\bar l^{c}$).
The supersymmetric minima of the scalar potential 
lie on the D flat direction $\phi=\bar{\phi}^*$ 
($l^c=\bar l^{c*}$) at $\langle S\rangle = 0~,
~|\langle\phi\rangle|=|\langle\bar{\phi}\rangle|=M$ 
($|\langle l^c\rangle|=|\langle\bar l^{c}\rangle|=M$).

\par
Hybrid inflation \cite{hybrid} is `naturally' and 
automatically realized \cite{lyth,dss} in such 
supersymmetric schemes. The scalar potential possesses 
a built-in inflationary trajectory at $|S|>M$, 
$\phi=\bar{\phi}=0$ ($l^c =\bar l^{c}=0$) with 
a constant tree-level potential energy density 
$\kappa^{2}M^{4}$ which causes the exponential 
expansion of the universe. Moreover, since 
this constant energy density breaks supersymmetry and 
produces mass splitting in the supermultiplets $\phi$, 
$\bar{\phi}$ ($l^c$, $\bar l^{c}$), there 
are important radiative corrections \cite{dss} which 
provide a slope along the inflationary trajectory 
necessary for driving the inflaton towards the vacua.
At one-loop, the cosmic microwave quadrupole anisotropy, 
in the $G_{B-L}$ model, is given \cite{deg} by  
$$
\left(\frac{\delta T}{T}\right)_{Q}\approx 8
\pi\left(\frac{N_{Q}}{45}\right)^{1/2}
\left(\frac{M}{M_{P}}\right)^{2}
$$
\begin{equation}
x_{Q}^{-1}y_{Q}^{-1}
\Lambda (x_{Q}^{2})^{-1}~,
\label{quadrupole}
\end{equation} 
where $N_Q \approx 50-60$ denotes the number of 
e-foldings experienced by our present horizon size during 
inflation, $M_{P}\approx 1.22\times 10^{19}~{\rm{GeV}}$ 
is the Planck scale and
\begin{equation}
\Lambda (z)=
(z-1)\ln (1-z^{-1})+(z+1)\ln (1+z^{-1})~.
\label{lambda}
\end{equation} 
Also, 
\begin{equation}
y_Q^2=\int_{1}^{x_Q^{2}}\frac{dz}{z}
\Lambda(z)^{-1}~,~y_Q \geq 0~,
\label{yQ}
\end{equation}
with $x_{Q}=|S_{Q}|/M$ ($x_{Q}\geq 1$), $S_{Q}$ 
being the value of the scalar field $S$ when our present 
horizon scale crossed outside the inflationary horizon. 
The superpotential parameter $\kappa$, in the $G_{B-L}$ 
model, can be evaluated \cite{deg} from
\begin{equation} 
\kappa \approx \frac{8\pi ^{3/2}}
{\sqrt{N_{Q}}}~y_{Q}~\frac{M}{M_{P}}~\cdot
\label{kappa}
\end{equation}
Note that, in the $G_{LR}$ case, the right hand sides of 
Eqs.(\ref{quadrupole}) and (\ref{kappa}) should be divided 
by $\sqrt{2}$ \cite{hier}. This is due to the fact that the 
replacement of $\phi,~\bar{\phi}$ by $l^c$, $\bar l^{c}$ 
doubles the one-loop contribution to the effective 
`inflationary' potential.

\par
The $\mu$ term can be generated \cite{dls} by adding the 
superpotential coupling
\begin{equation} 
\delta W=\lambda S\epsilon^{ij}H^{(1)}_{i}H^{(2)}_{j}
=\lambda S H^{2}~~~(\lambda>0)~,
\label{higgs}
\end{equation} 
where $H=(H^{(1)}, H^{(2)})$ is the electroweak higgs 
superfield belonging, in the $G_{LR}$ case, to a bidoublet 
$(2,2)_{0}$ representation of $SU(2)_L\times SU(2)_R
\times U(1)_{B-L}$~. After gravity-mediated supersymmetry 
breaking, $S$ develops \cite{dls} a vacuum expectation 
value (vev) $\langle S\rangle\approx -m_{3/2}/\kappa$~, 
where $m_{3/2}\sim (0.1-1)\ {\rm TeV}$ is the gravitino 
mass, and generates a $\mu$ term with $\mu=\lambda 
\langle S\rangle\approx -(\lambda/\kappa)m_{3/2}$~. 

\par
This particular solution of the $\mu$ problem is 
\cite{deg,atmo}, however, not totally satisfactory 
since it requires the presence of `unnaturally' small 
coupling constants 
($\kappa\stackrel{_{<}}{_{\sim }}\times 10^{-5}$). 
This is due to the fact that the inflaton system 
decays predominantly into electroweak higgs superfields
via the renormalizable superpotential coupling in 
Eq.(\ref{higgs}). 
The gravitino constraint \cite{gravitino} on the `reheat' 
temperature then severely restricts the corresponding 
dimensionless coupling constant and, consequently, the 
parameter $\kappa$. Moreover, for hierarchical neutrino 
masses from the seesaw mechanism, the requirement of 
maximal $\nu_{\mu}-\nu_{\tau}$ mixing from the 
SuperKamiokande experiment \cite{superk} further reduces 
\cite{atmo} $\kappa$ to become of order $10^{-6}$.

\par
We adopt an alternative solution of the $\mu$ problem 
constructed \cite{kn} by coupling the electroweak 
higgses to superfields causing the breaking of the 
Peccei-Quinn symmetry ($U(1)_{PQ}$). We introduce two 
extra gauge singlet left handed superfields $N$ and 
$\bar{N}$ with $PQ$ charges -1 and 1 respectively. The 
relevant superpotential couplings are $\lambda N^{2} 
\bar{N}^2/2m_{P}$ ($m_{P}\equiv M_{P}/\sqrt{8\pi} 
\approx 2.44\times 10^{18}~{\rm{GeV}})$ and 
$N^{2}H^{(1)}H^{(2)}$ (or $N^{2}H^{2}$). After 
gravity-mediated supersymmetry breaking, the scalar 
potential generated by $N^{2}\bar{N}^2$ is \cite{rsym}
$$
\left(m_{3/2}^2+\lambda^2\left|
\frac{N\bar{N}}{m_P}\right|^2\right)
\left[(|N|-|\bar{N}|)^2+2|N||\bar{N}|\right]
$$
\begin{equation}
+|A|m_{3/2}
\lambda\frac{|N\bar{N}|^2}{m_P}{\rm{cos}}
(\epsilon+2\theta+2\bar{\theta})~,
\label{eq:npotential}
\end{equation}
where $\epsilon,~\theta,~\bar{\theta}$ are the phases of 
$A,~N,~\bar{N}$. Minimization of this 
potential then requires $\epsilon+2\theta+2\bar{\theta}
=\pi$, $|\langle N\rangle|=|\langle\bar{N}\rangle|$ and, 
for $|A|>4$,
$$
|\langle N\rangle|=(m_{3/2}m_P)^{\frac{1}{2}}
\left(\frac{|A|+(|A|^2-12)^{\frac{1}{2}}}
{6\lambda}\right)^{\frac{1}{2}}
$$ 
\begin{equation}
\sim (m_{3/2}m_{P})^{\frac{1}{2}}
\sim 10^{11}~{\rm{GeV}}~.
\label{eq:solution}
\end{equation}
This scale is identified with the symmetry breaking 
scale $f_{a}$ of $U(1)_{PQ}$~. Substitution of 
$\langle N \rangle$ in the superpotential coupling 
$N^{2}H^{(1)}H^{(2)}$ (or $N^{2}H^{2}$) generates 
a $\mu$ parameter of order $m_{3/2}$.

\par
This resolution of the $\mu$ problem avoids the direct 
coupling of the inflaton system $S$, $\phi$, 
$\bar{\phi}$ (or $S$, $l^c$, $\bar l^{c}$) 
to the electroweak higgses. Thus, the inflaton does 
not predominantly decay into higgses via renormalizable 
couplings as in the previous case. It decays to 
$SU(2)_{L}$ triplets $T_a$, $\bar{T}_a$ (or right 
handed neutrino superfields $\nu^{c}$) via 
nonrenormalizable interactions, which are `naturally' 
suppressed by $m_{P}^{-1}$ (see below) . The gravitino 
constraint can be satisfied with more `natural' values 
of the dimensionless parameters \cite{deg,hier}.

\par
We now proceed to the detailed description of the two 
models \cite{deg,hier}. The superpotential $W$ contains, 
in addition to the terms in Eq.(\ref{W}), the following 
couplings in the two cases:
\begin{equation}
\begin{array}{rlc}
G_{B-L}:H^{(1)}QU^c,H^{(2)}QD^c,H^{(2)}LE^c,
N^{2} \bar{N}^2,
\\
N^{2}H^{(1)}H^{(2)},TLL,\bar{T} H^{(1)} H^{(1)},
\bar{\phi}^{2}T\bar{T};~~~~~
\end{array}
\label{couplbl}
\end{equation}
\begin{equation}
G_{LR}:HQQ^c,HLL^c,N^{2} \bar{N}^2,
N^{2}H^{2},\bar l^{c}\bar l^{c}L^{c}L^{c}.
\label{coupllr}
\end{equation}
Here $Q_i$ and $L_i$ denote the $SU(2)_{L}$ doublet left 
handed quark and lepton superfields, whereas the superfields
$Q_i^c=(U_i^c, D_i^c)$ and  $L^c_i=(\nu^c_i, E^c_i)$ 
are the $SU(2)_{L}$ singlet ($SU(2)_{R}$ doublet) 
antiquarks and antileptons ($i$=1,2,3 is the family index).
Of course, the right handed neutrino superfields $\nu^c_i$
are absent in the $G_{B-L}$ case, where two pairs of 
$SU(2)_{L}$ triplets $T_a$, $\bar{T}_a$ ($a=1,2$) with 
$Y=1,-1$ and $B-L=2,0$ respectively are included.

\par
The continuous global symmetries of the superpotential 
are $U(1)_B$ (and, thus, $U(1)_L$) with the extra 
chiral superfields $S$, $\phi$, $\bar{\phi}$, $N$, 
$\bar{N}$, $T$, $\bar{T}$, $l^{c}$, $\bar l^{c}$ 
carrying zero baryon number, an anomalous Peccei-Quinn 
symmetry $U(1)_{PQ}~$, and a non-anomalous R-symmetry 
$U(1)_{R}~$. The $PQ$ charges of the superfields, in 
the two cases, are as follows:
\begin{equation}
\begin{array}{rlc}
G_{B-L}:H^{(1)}(1),H^{(2)}(1),L(-1),E^c(0),
\\
Q(-1),U^c(0),D^c(0),S(0),\phi(0),\bar{\phi}(0),~
\\
N(-1),\bar{N}(1),T(2),\bar{T}(-2);~~~~~~~~
\end{array}
\label{pqbl}
\end{equation}
\begin{equation}
\begin{array}{rlc}
G_{LR}:H(1),L(-1),L^c(0),Q(-1),Q^c(0),
\\
S(0),l^{c}(0),\bar l^{c}(0),N(-1),\bar{N}(1).~~~~~~
\end{array}
\label{pqlr}
\end{equation}
The $R$ charges of the superfields are ($W$ carries one 
unit of $R$ charge):
\begin{equation}
\begin{array}{rlc}
G_{B-L}:H^{(1)}(0),H^{(2)}(0),L(1/2),~~~~
\\
E^c(1/2),Q(1/2),U^c(1/2),D^c(1/2),S(1),
\\
\phi(0),\bar{\phi}(0),N(1/2),\bar{N}(0),T(0),\bar{T}(1);~~
\end{array}
\label{rbl}
\end{equation}
\begin{equation}
\begin{array}{rlc}
G_{LR}:H(0),L(1/2),L^c(1/2),Q(1/2),~~
\\
Q^c(1/2),S(1),l^{c}(0),\bar l^{c}(0),N(1/2),\bar{N}(0).
\end{array}
\label{rlr}
\end{equation}

\par
Note that $U(1)_{B}$ (and, thus, $U(1)_{L}$) is 
automatically implied by $U(1)_{R}$ even if all 
possible nonrenormalizable terms are included. This 
is due to the fact that the $R$ charges of the 
products of any three color (anti)triplets exceed 
unity and cannot be compensated since there are no 
negative $R$ charges available. 

\par 
To avoid undesirable mixing of $L$ 's with the higgs 
$H^{(2)}$ or $\bar l^{c}$ via the allowed superpotential 
couplings $N\bar{N}LH^{(1)}\phi$, $N\bar{N}LHl^{c}$, 
$N\bar{N}L^{c}\bar l^{c}$, we impose an extra discrete 
$Z_2$ symmetry (`lepton parity') under which $L$, 
$L^c=(\nu^c, E^c)$ change sign. This symmetry is 
equivalent to `matter parity' (under which $L$, 
$L^c=(\nu^c, E^c)$, $Q$, $Q^c=(U^c, D^c)$ change sign), 
since `baryon parity' (under which $Q$, $Q^c=(U^c, D^c)$ 
change sign) is also present being a subgroup of $U(1)_B$~. 

\par 
The only superpotential terms which are permitted by the 
global symmetries $U(1)_{R}$~, $U(1)_{PQ}$ and `lepton 
parity' are the ones in Eqs.(\ref{W}) and (\ref{couplbl}) 
(or (\ref{coupllr})) as well as $LLl^cl^c\bar{N}^2 l^c
\bar l^c$ and $LLl^cl^cHH$, in the $G_{LR}$ case, modulo 
arbitrary multiplications by nonnegative powers of the 
combination $\phi\bar{\phi}$ (or $l^c\bar l^c$). The 
vevs of $\phi$, $\bar{\phi}$ (or $l^c$, $\bar l^c$) and 
$N$, $\bar{N}$ leave unbroken only the symmetries $G_{S}$, 
$U(1)_{B}$ and `matter parity'.

\par
We will first concentrate on the model based on the 
$G_{B-L}$ gauge group and discuss in some detail 
neutrino mass generation and baryogenesis in its 
context. After $B-L$ (and lepton number) breaking at 
the superheavy scale $M$, the last term in 
Eq.(\ref{couplbl}) generates intermediate scale 
masses for the $SU(2)_L$ triplet superfields $T_a$, 
$\bar{T}_a$ ($a$=1,2).  These masses can be taken 
positive and diagonal by appropriate transformations
and are given by $M_a=\gamma_a M^{2}/m_{P}$ 
($\gamma_a$ ($a$=1,2) are the dimensionless coupling 
constants of the terms $m_{P}^{-1}\bar{\phi}^{2}T_{a}
\bar{T}_{a}$). Also, after the electroweak breaking, the 
last two terms in Eq.(\ref{couplbl}) give rise to terms 
linear with respect to $T_a$ 's in the scalar potential. 
The $T_a$ 's then acquire vevs given by 
$\langle T_a\rangle=\beta_{a}\langle H^{(1)}
\rangle^{2}/M_{a}\sim M_{W}^{2}/M\ll M_{W}$~,
with $\beta_{a}$ being the coupling 
constant of the term $\bar{T}_aH^{(1)} H^{(1)}$. 
These vevs violate lepton number and, substituted to 
the coupling $TLL$ in Eq.(\ref{couplbl}), generate 
nonzero masses for light neutrinos. The neutrino mass 
matrix can be diagonalized by a suitable 
`Kobayashi-Maskawa' rotation in its standard form 
(involving three angles and a CP violating phase) and 
the complex eigenvalues can be written as
\begin{equation} 
m_i=\sum_{a=1,2}\alpha_{ai}\beta_{a}
\frac{\langle H^{(1)}\rangle^{2}}{M_a}~,
\label{mass}
\end{equation}
where $\alpha_{ai}$ are the (complex) eigenvalues of 
the complex symmetric coupling constant matrix of the 
term $T_{a}L_{i}L_{j}$. Note that the $m_i$ 's, being 
in general complex, carry two extra CP violating phases 
(an overall phase factor is irrelevant) which appear 
in some processes like neutrinoless double-beta decay.

\par
We take degenerate light neutrino masses, which are not
inconceivable here as in the seesaw case. Neutrinos can 
then provide the hot dark matter of the universe needed
for explaining \cite{structure} its large scale structure 
for $\Lambda=0$ without any conflict with 
atmospheric/solar neutrino oscillations even within a 
three neutrino scheme.

\par
For definiteness, we can take the model of neutrino 
masses and mixing discussed in Ref.\cite{gg}, although 
the precise values of mixing angles and square-mass 
differences are not relevant for our discussion.
This scheme has almost degenerate neutrino masses and 
employs the bimaximal neutrino mixing \cite{bimaximal} 
which is consistent with the vacuum oscillation explanation 
\cite{vacuum} of the solar neutrino puzzle. Moreover, 
all three neutrino masses are real, but the CP parity of 
one of them (say the second one) is opposite to the CP 
parities of the other two. This is important for satisfying 
the experimental constraints \cite{beta} from neutrinoless 
double-beta decay. The neutrino scheme of Ref.\cite{gg} can 
be obtained in our model provided the coupling constants 
$\alpha_{ai}$ ($a$=1,2; $i$=1,2,3) satisfy the relations 
$\alpha_{a1}=-\alpha_{a2}=\alpha_{a3}\equiv\alpha_{a}$ 
to a very good approximation and this will be the only 
information we will use from this scheme.

\par
We now turn to the discussion of the decay of the inflaton, 
which consists of the two complex scalar fields $S$ and 
$\theta=(\delta\phi+\delta\bar{\phi})/\sqrt{2}$, where 
$\delta\phi=\phi-M$, $\delta \bar{\phi}=\bar{\phi}-M$, 
with mass $m_{infl}=\sqrt{2}\kappa M$. The scalar $\theta$ 
($S$) can decay into a pair of fermionic (bosonic) $T_a$, 
$\bar{T}_a$ 's as one easily deduces from the last coupling 
in Eq.(\ref{couplbl}) and the coupling $\kappa S\phi
\bar{\phi}$ in Eq.(\ref{W}). The decay width is the same 
for both scalars and equals
\begin{equation} 
\Gamma=\frac{3}{8\pi}~\gamma_{a}^{2}
\left(\frac{M}{m_P}\right)^{2}m_{infl}~.
\label{width}
\end{equation}
Of course, decay of the inflaton into $T_a$, $\bar{T}_a$ 
is possible provided that the corresponding triplet mass 
$M_a \leq m_{infl}/2$. The gravitino constraint 
\cite{gravitino} on the `reheat' temperature, $T_r~$, 
then implies strong bounds on the $M_a$ 's which satisfy 
this inequality. Consequently, the corresponding 
dimensionless coupling constants, $\gamma_{a}$~, are 
restricted to be quite small.

\par
To minimize the number of small couplings, we then take 
$M_2 < m_{infl}/2\leq M_1=M^{2}/m_{P}$ ($\gamma_{1}=1$) 
so that the inflaton decays into only one (the lightest) 
triplet pair with mass $M_2$. Using Eq.(\ref{kappa}), 
the requirement $m_{infl}/2\leq M_1$ becomes $y_{Q}\leq 
\sqrt{N_{Q}}/2\pi \approx 1.2$, for $N_{Q}=56$, and 
Eq.(\ref{yQ}) gives $x_{Q}\stackrel{_{<}}{_{\sim }} 
1.6$. As an example, we 
choose $x_{Q}=1.2$ which corresponds to $y_{Q}=0.61$ 
(see Eq.(\ref{yQ})). Eqs.(\ref{quadrupole}), 
(\ref{kappa}) with $(\delta T/T)_{Q}\approx 6.6\times 
10^{-6}$ from the cosmic background explorer (COBE) 
\cite{cobe} then give $M \approx 4.43
\times 10^{15}~{\rm{GeV}}$ and $\kappa \approx 1.32
\times 10^{-3}$. Also, the inflaton mass is $m_{infl} 
\approx 8.27\times 10^{12}~{\rm{GeV}}$, the $SU(2)_{L}$ 
triplet masses are $M_{1}\approx 8.04\times 10^{12}~{\rm{GeV}}$, 
$M_{2} \approx 8.04\gamma_{2}\times 10^{12}~{\rm{GeV}}$, 
and the `reheat' temperature is $T_{r} \approx 8.19
\gamma_{2}\times 10^{11}~{\rm{GeV}}$. The gravitino 
constraint \cite{gravitino} for MSSM spectrum ($T_r\approx 
(1/7)(\Gamma M_{P})^{1/2}\stackrel{_{<}}{_{\sim }}
10^9$ GeV) then implies $\gamma_{2}
\stackrel{_{<}}{_{\sim }}1.11\times 10^{-3}$.

\par
In this scheme, baryon number is violated only  by `tiny' 
nonperturbative $SU(2)_L$ instanton effects. So the only 
way to produce the observed BAU is to first generate a 
primordial lepton asymmetry \cite{leptogenesis} which 
is then partially converted to baryon asymmetry by  
sphalerons. The primordial lepton 
asymmetry is produced via the decay of the superfields 
$T_{2}$, $\bar{T}_{2}$ which emerge as decay products 
of the inflaton. This mechanism for leptogenesis has been 
discussed in Refs.\cite{rsym,sarkar}. The $SU(2)_L$ 
triplet superfields decay either to a pair of 
$L_i$ 's or to a pair of $H^{(1)}$ 's. The relevant 
one-loop diagrams are \cite{sarkar} of the self-energy 
type \cite{covi} with a s-channel exchange of $T_{1}$, 
$\bar{T}_{1}$. The resulting lepton asymmetry is 
\cite{sarkar}
$$
\frac {n_{L}}{s} \approx -1.33~\frac{3}{8 \pi}~
\frac {T_{r}}{m_{infl}}
$$ 
\begin{equation}
\frac{M_{1}M_{2}}{M_{1}^{2}-M_{2}^{2}}~
\frac{{\rm{Im}}(\beta_{1}^{*}\beta_{2}
{\rm{Tr}}(\hat{\alpha}_{1}^{\dagger}
\hat{\alpha}_{2}))}
{\rm{Tr}(\hat{\alpha}_{2}^{\dagger}
\hat{\alpha}_{2})
+\beta_{2}^{*}\beta_{2}}~,
\label{asymmetry}
\end{equation}
where $\hat{\alpha}_{a}={\rm{diag}}(\alpha_{a},
-\alpha_{a}, \alpha_{a})$. Note that this formula 
holds provided \cite{pilaftsis} the decay width of $T_{1}$, 
$\bar{T}_{1}$ is much smaller than 
$(M_{1}^{2}-M_{2}^{2})/M_{2}$, which is well satisfied 
here since $M_{2}\ll M_{1}$. For MSSM spectrum, the 
observed BAU is given \cite{ibanez} by $n_{B}/s=
-(28/79)(n_{L}/s)$. It is important to ensure that the 
primordial lepton asymmetry is not erased by lepton number 
violating $2\rightarrow 2$ scattering processes at all 
temperatures between $T_r$ and 100 GeV. This gives 
\cite{ibanez} $m_{\nu_{\tau}}\stackrel{_<}{_\sim} 
10~{\rm{eV}}$ which is readily satisfied.

\par
The parameters $\alpha_{a}$, $\beta_{a}$, $\gamma_{a}$
($a$=1,2) are constrained by the requirement that the hot 
dark matter of the universe consists of light neutrinos. 
We take the `relative' density of hot dark matter 
$\Omega_{HDM} \approx 0.2$, which is favored
by the structure formation in cold plus hot dark matter 
models \cite{structure} with $\Lambda=0$, and 
$h \approx 0.5$, where $h$ 
is the present value of the Hubble parameter in units of 
$100~\rm{km}~\rm{sec}^{-1}~\rm{Mpc}^{-1}$. The 
common mass of the three light neutrinos is then about 
$1.5~{\rm{eV}}$ and Eq.(\ref{mass}) gives the
constraint
\begin{equation} 
\left|\sum_{a=1,2}\frac{\alpha_{a}\beta_{a}}
{\gamma_{a}}\right| \approx\left(
\frac{M}{7.02\times 10^{15}~{\rm{GeV}}}\right)^{2}
\equiv \xi~,
\label{constraint}
\end{equation}
where $|\langle H^{(1)}\rangle|$ was taken $\approx 
174~{\rm{GeV}}$. Lepton asymmetry is maximized, under 
this constraint, for $|\alpha_{1}\beta_{1}/\gamma_{1}|
=|\alpha_{2}\beta_{2}/\gamma_{2}|\equiv \delta$ and
$\sqrt{3}|\alpha_{2}|=|\beta_{2}|$. Substituting
$T_r\approx (1/7)(\Gamma M_{P})^{1/2}$ with $\Gamma$
from Eq.(\ref{width}), Eq.(\ref{asymmetry}) gives
\begin{equation}
\left|\frac {n_{L}}{s}\right| 
\stackrel{_{<}}{_{\sim }}\frac{0.107}{\pi}~
\frac{M}{\sqrt{m_{infl}M_{P}}}~
\gamma_{1}\gamma_{2}^{2}~\xi\left(1-
\frac{\xi^{2}}{4\delta^{2}}\right)^{1/2}
\cdot
\label{maximum}
\end{equation}
which is further maximized at $\alpha_{1}=\beta_{1}=1$. 
This gives $\delta=1$ (for $\gamma_{1}=1$). 
For $x_{Q}=1.2$, $\xi \approx 0.4$ and 
the maximal lepton asymmetry becomes $\approx 5.86~
\gamma_{2}^{2}\times 10^{-3}$. The low deuterium 
abundance constraint \cite{deuterium} on the BAU , 
$\Omega _{B}h^{2}\approx 0.019$, 
can then be satisfied provided $\gamma_{2}
\stackrel{_{>}}{_{\sim }} 1.88\times 10^{-4}$. So, for 
$\gamma_{2}$ in the range $1.9\times 10^{-4}-1.1\times 
10^{-3}$, both the gravitino and baryogenesis restrictions 
can be met. We see that, in the $G_{B-L}$ model, the 
required values  of the relevant coupling constants 
$\kappa$ and $\gamma_{2}$ are more or less `natural' 
($\sim 10^{-3}$). 

\par
We now turn to the discussion of the second model based on 
the left-right symmetric gauge group $G_{LR}$. After 
$B-L$ breaking by $\langle l^c\rangle$, $\langle
\bar l^{c}\rangle$, the last term in Eq.(\ref{coupllr}) 
generates intermediate scale masses for the right handed 
neutrino superfields $\nu^{c}_i$ ($i$=1,2,3). The 
dimensionless coupling constant matrix of this term can be 
made diagonal with positive entries $\gamma_i$ 
($i$=1,2,3) by a rotation on $\nu^{c}_i$ 's. The right 
handed neutrino mass eigenvalues are then $M_i=2\gamma_i 
M^{2}/m_{P}$ (with $\langle l^{c}\rangle$, 
$\langle \bar l^{c} \rangle$ taken positive by a $B-L$ 
transformation). 

\par
The light neutrino masses are generated via the seesaw 
mechanism and, therefore, cannot be `naturally' 
degenerate. We will, thus, assume hierarchical 
light neutrino masses. Analysis \cite{giunti} of the 
CHOOZ experiment \cite{chooz} shows that the oscillations 
of solar and atmospheric neutrinos decouple. This fact 
allows us to concentrate on the two heaviest families 
ignoring the first one. We will denote the two positive 
eigenvalues of the light neutrino mass matrix by $m_{2}$ 
(=$m_{\nu _{\mu }}$), $m_{3}$ (=$m_{\nu _{\tau }}$). 
We take $m_{\nu_{\mu}}\approx 2.6\times 
10^{-3}~\rm{eV}$ which is the central value of the 
$\mu$-neutrino mass coming from the small angle MSW 
resolution of the solar neutrino problem \cite{smirnov}. 
The $\tau$-neutrino mass is taken to be 
$m_{\nu _{\tau }}\approx 7\times 10^{-2}~\rm{eV}$ 
which is the central value implied by SuperKamiokande 
\cite{superk}. 

\par
The determinant and the trace invariance of the light 
neutrino mass matrix imply \cite{neu} two constraints 
on the (asymptotic) parameters: 
\begin{equation}
m_{2}m_{3}\ =\ \frac{\left( m_{2}^{D}m_{3}^{D}
\right) ^{2}}{M_{2}\ M_{3}}~,
\label{determinant}
\end{equation}
\begin{eqnarray*}
m_{2}\,^{2}+m_{3}\,^{2}\ =
\end{eqnarray*}
\begin{eqnarray*}
\frac{\left( m_{2}^{D}\,\,^{2}{\rm c}^{2}+
m_{3}^{D}\,^{2}{\rm s}^{2}\right)^{2}}{M_{2}\,^{2}}+
\ \frac{\left( m_{3}^{D}\,^{2}{\rm c}^{2}+
m_{2}^{D}\,^{2}{\rm s}^{2}\right)^{2}}{M_{3}\,^{2}}
\end{eqnarray*}
\begin{equation}
+\ \frac{2(m_{3}^{D}\,^{2}-m_{2}^{D}\,^{2})^{2}
{\rm c}^{2}{\rm s}^{2}\,{\cos 2\delta }}
{M_{2}\,M_{3}}~\cdot
\label{trace} 
\end{equation}
Here, $m_{2,3}^{D}$ ($m_{2}^{D}\leq m_{3}^{D}$) are 
the `Dirac' neutrino masses considered diagonal, and
${\rm c}=\cos \theta ,\ {\rm s}=\sin \theta $ 
with $\theta$ and $\delta$ being the rotation angle and 
phase which diagonalize the Majorana mass matrix of the 
right handed neutrinos.

\par
The $\nu_{\mu}-\nu_{\tau}$ mixing angle 
$\theta _{\mu\tau}$ lies in the range
\begin{equation}
|\,\varphi -\theta ^{D}|\leq \theta _{\mu\tau}\leq
\varphi +\theta ^{D},\ {\rm {for}\ \varphi +
\theta }^{D}\leq \ \pi /2~,
\label{mixing}
\end{equation}
where $\varphi$ is the rotation angle diagonalizing 
the light neutrino mass matrix and $\theta ^{D}$ the 
`Dirac' (unphysical) mixing angle defined in the absence 
of right handed neutrino Majorana masses \cite{neu}.

\par
We will now discuss the `reheating' process in the $G_{LR}$ 
model. The inflaton again consists of the two complex 
scalar fields $S$ and $\theta$ ($\phi$, $\bar{\phi}$ are 
now the neutral components of $l^{c}$, $\bar{l^{c}}$). In 
this case, however, the scalar $S$ ($\theta$) decays into 
a pair of bosonic (fermionic) $\nu^{c}_{i}$ 's via the last 
coupling in Eq.(\ref{coupllr}) and $\kappa Sl^{c}
\bar{l^{c}}$ with a decay width
\begin{equation} 
\Gamma=\frac{1}{8\pi}~\left(\frac{M_{i}}
{M}\right)^{2}m_{infl}~,
\label{widthlr}
\end{equation}
provided that $M_i < m_{infl}/2$. The gravitino constraint 
implies strong bounds on these $M_i$ 's and, consequently, on 
the corresponding $\gamma_{i}$ 's.

\FIGURE{\epsfig{figure=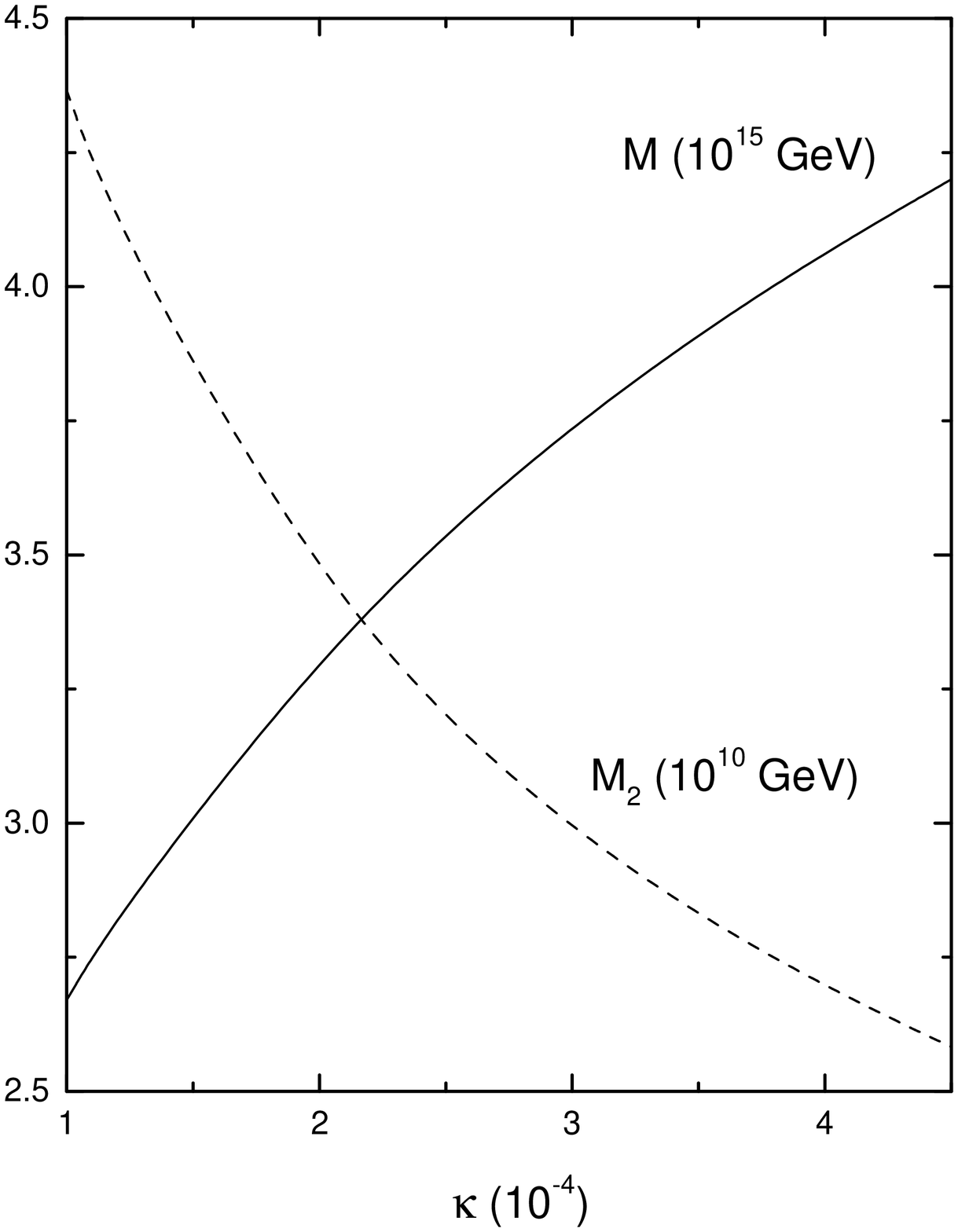,height=3.25in,angle=0}
\medskip
\caption{The mass scale $M$ (solid line) and the Majorana 
mass of the second heaviest right handed neutrino $M_{2}$ 
(dashed line) as functions of $\kappa$.
\label{M}}}

\par
We minimize the number of small couplings by taking 
$M_2 < m_{infl}/2\leq M_3=2M^{2}/m_{P}$ 
($\gamma_{3}=1$) so that the inflaton decays into only 
one (the second heaviest) right handed neutrino with mass 
$M_2$. The second inequality implies $y_{Q}\leq 
\sqrt{2N_{Q}}/\pi \approx 3.34$ (for $N_{Q}=55$) and, 
thus, $x_{Q}\stackrel{_{<}}{_{\sim }} 3.5$. 
The parameters $M$ and $\kappa$ are 
calculated for each value of $x_{Q}$ in this range. 
Eliminating $x_{Q}$, we obtain $M$ as a function of 
$\kappa$ depicted in Fig.\ref{M}. The inflaton mass 
$m_{infl}$ and the heaviest right handed neutrino mass 
$M_{3}$ are readily evaluated. The mass of the second 
heaviest right handed neutrino $M_{2}$ is restricted by 
the gravitino constraint. We take it to be equal to its 
maximal allowed value in order to maximize $\gamma_{2}$. 
The value of $M_{2}$ is also depicted in Fig.\ref{M}.

\par
Baryogenesis proceeds through a primordial leptogenesis 
\cite{leptogenesis} in this model too. The lepton 
asymmetry, however, is now produced through the decay of the 
superfield $\nu^{c}_{2}$ which emerges as decay product 
of the inflaton. This mechanism for leptogenesis has been 
discussed in Ref.\cite{leptogenesis}. The $\nu^{c}_{2}$ 
superfield decays into electroweak higgs and (anti)lepton 
superfields. The relevant one-loop diagrams are both of the 
vertex and self-energy type \cite{covi} with an exchange 
of $\nu^{c}_{3}$. The resulting lepton asymmetry 
is \cite{neu}
\begin{eqnarray*}
\frac{n_{L}}{s}\approx 1.33~\frac{9T_{r}}
{16\pi m_{infl}}~\frac{M_2}{M_3}
\end{eqnarray*}
\begin{equation}
\frac{{\rm c}^{2}{\rm s}^{2}\ 
\sin 2\delta \ 
(m_{3}^{D}\,^{2}-m_{2}^{D}\,^{2})^{2}}
{|\langle H^{(1)}\rangle|^{2}~(m_{3}^{D}\,^{2}\ 
{\rm s}^{2}\ +
\ m_{2}^{D}\,^{2}{\rm \ c^{2}})}~\cdot
\label{leptonasym}
\end{equation}
Note that this formula holds \cite{pilaftsis} provided 
that $M_{2}\ll M_{3}$ and the decay width of 
$\nu^{c}_{3}$ is much smaller than 
$(M_{3}^{2}-M_{2}^{2})/M_{2}$, 
and both conditions are well satisfied here. The `dangerous' 
lepton number violating processes are well out of equilibrium 
in this case too.

\FIGURE{\epsfig{figure=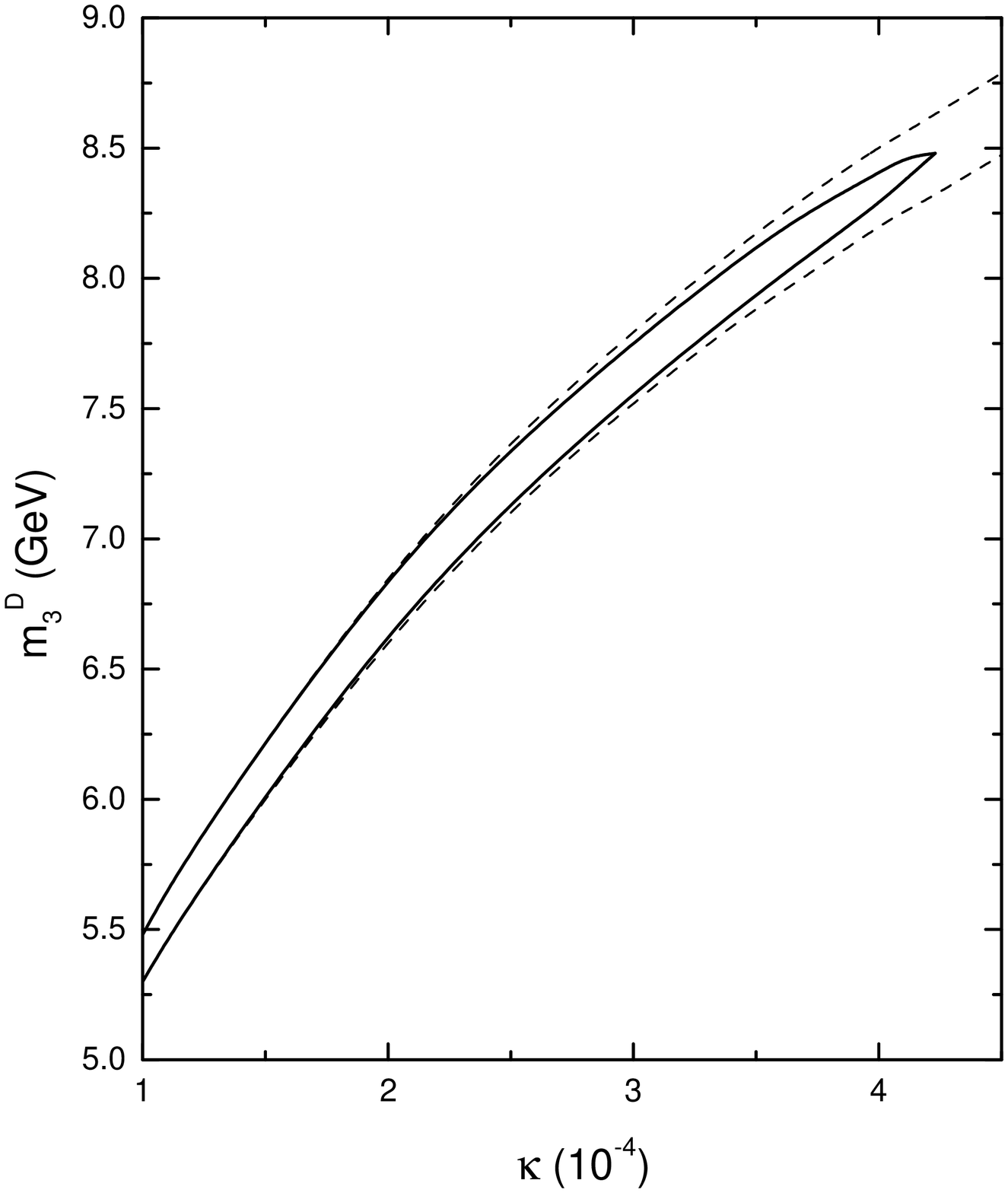,height=3in,angle=0}
\medskip
\caption{The area (bounded by the dashed lines) on the 
$\kappa-m^{D}_{3}$ plane consistent with maximal 
$\nu_{\mu}-\nu_{\tau}$ mixing and the gravitino 
constraint. Along the thick solid line the low deuterium 
abundance constraint on the BAU is also satisfied.
\label{mD3}}}

\par
For definiteness, we assume that the $\nu_{\mu}-
\nu_{\tau}$ mixing is about maximal 
($\theta _{\mu\tau}\approx\pi/4$) in accordance with 
the recent SuperKamiokande data \cite{superk}. We will 
also make the plausible assumption that the `Dirac' mixing 
angle $\theta^{D}$ is negligible 
($\theta^{D}\approx 0$). Under these circumstances, the 
rotation angle  $\varphi\approx\pi/4$. Using the 
`determinant' and `trace' constraints in 
Eqs.(\ref{determinant}) and (\ref{trace}) and 
diagonalizing the light neutrino mass matrix, we can 
determine the range of $m^{D}_{3}$ which allows maximal 
$\nu_{\mu}-\nu_{\tau}$ mixing for each value of $\kappa$.
These ranges are depicted in Fig.\ref{mD3} for all 
relevant values of $\kappa$ and constitute the area in 
the $\kappa-m^{D}_{3}$ plane consistent with maximal 
mixing. For each allowed pair $\kappa$, $m^{D}_{3}$, 
the value of the phase $\delta$ leading to maximal 
mixing can be determined from the `trace' condition. 
The corresponding lepton asymmetry is then found from
Eq.(\ref{leptonasym}). The line consistent with the
low deuterium abundance constraint \cite{deuterium} on 
the BAU ($\Omega_{B}h^2\approx 0.019$) is also 
depicted in Fig.\ref{mD3}. We
see that the required values of $\kappa$ ($\stackrel{_<}
{_\sim}4.2\times 10^{-4}$), although somewhat small, are
much more `natural' than the ones encountered in previous 
models \cite{atmo} that solved the $\mu$ problem 
and achieved maximal $\nu_{\mu}-\nu_{\tau}$ mixing. 
For these values of $\kappa$, $\gamma_{2}
\stackrel{_>}{_\sim}1.9\times 10^{-3}$ which is quite 
satisfactory.

\par
In conclusion, we have presented two moderate extensions of
MSSM based on the gauge groups $G_{B-L}$ and $G_{LR}$. In
the $G_{B-L}$ case, neutrinos acquire degenerate masses, 
thereby providing the hot dark matter in the universe needed
for explaining its large scale structure especially for zero 
cosmological constant. In the case of the left-right 
symmetric gauge group $G_{LR}~$, neutrino masses are 
generated via the seesaw mechanism and are taken hierarchical. 
The recent SuperKamiokande restrictions on $m_{\nu_{\tau}}$ 
and $\nu_{\mu}-\nu_{\tau}$ mixing can be accommodated, in 
this model, with $m_{\nu_{\mu}}$ from the small angle 
MSW resolution of the solar neutrino puzzle. Hybrid inflation 
is `naturally' realized and the $\mu$ problem is easily 
resolved by a Peccei-Quinn symmetry in both models. Also, 
the BAU is generated through a primordial leptogenesis and 
the gravitino and baryogenesis constraints are easily met 
with more or less `natural' values of the parameters. 

\acknowledgments

\par
This work is supported by E.U. under TMR contract 
No. ERBFMRX--CT96--0090.


\begin{thebibliography}{99}

\bibitem{superk} T. Kajiata, talk given at the XVIIIth 
International Conference on Neutrino Physics and 
Astrophysics (Neutrino'98), Takayama, Japan, 4-9 June, 1998.

\bibitem{structure} Q. Shafi and F. W. Stecker, 
\prl{53}{1984}{1292}. For a recent review and other 
references see Q. Shafi and R. K. Schaefer, hep-ph/9612478.

\bibitem{deg} G. Lazarides, \plb{452}{1999}{227}.

\bibitem{lss} G. Lazarides, R. Schaefer and Q. Shafi, 
\prd{56}{1997}{1324}.

\bibitem{hier} G. Lazarides and N. Vlachos, 
hep-ph/9903511 (to appear in {\it{Phys. Lett. {\bf B}}}).

\bibitem{triplet} G. Lazarides, Q. Shafi and C. Wetterich, 
\npb{181}{1981}{287}; C. Wetterich, \npb{187}{1981}{343}; 
R. N. Mohapatra and G. Senjanovic, \prd{23}{1981}{165}; 
R. Holman, G. Lazarides and Q. Shafi, \prd{27}{1983}{995}.

\bibitem{rsym} G. Lazarides and Q. Shafi, 
\prd{58}{1998}{071702}.

\bibitem{lambda} Brian P. Schmidt {\it{et al.}}, 
\apj{507}{1998}{46}; 
Adam G. Riess {\it{et al.}}, \anj{116}{1998}{1009};
S. Perlmutter {\it{et al.}}, astro-ph/9812473; 
S. Perlmutter {\it{et al.}}, astro-ph/9812133 
(to appear in {\it{Astrophys. Journal}}).

\bibitem{lambdafit} L. Krauss and M. S. Turner, 
\grg{27}{1995}{1137}; J. P. Ostriker and P. J. Steinhardt, 
\n{377}{1995}{600}; A. R. Liddle {\it{et al.}}, 
\mnras{282}{1996}{281}.

\bibitem{primack} J. R. Primack and M. A. K. Gross, 
astro-ph/9810204.

\bibitem{hybrid} A. D. Linde, \pl{259}{1991}{38}; 
\pr{49}{1994}{748}.

\bibitem{lyth} E. J. Copeland, A. R. Liddle, D. H. Lyth, 
E. D. Stewart and D. Wands, \prd{49}{1994}{6410}. 

\bibitem{dss} G. Dvali, Q. Shafi and R. Schaefer, 
\prl{73}{1994}{1886}.

\bibitem{dls} G. Dvali, G. Lazarides and Q. Shafi, 
\plb{424}{1998}{259}.

\bibitem{atmo} G. Lazarides and N. D. Vlachos, 
\plb{441}{1998}{46}.

\bibitem{gravitino} M. Yu. Khlopov and A. D. Linde, 
\pl{138}{1984}{265}; J. Ellis, J. E. Kim and D. Nanopoulos, 
\pl{145}{1984}{181}.

\bibitem{kn} Jihn E. Kim and Hans P. Nilles, 
\pl{138}{1984}{150}.

\bibitem{gg} H. Georgi and S. L. Glashow, hep-ph/9808293.

\bibitem{bimaximal} V. Barger, S. Pasvaka, T. Weiler 
and K. Whisnant, hep-ph/9806387.

\bibitem{vacuum} V. Barger, R. J. N. Phillips and 
K. Whisnant, \prd{24}{1981}{538}; S. L. Glashow and 
L. M. Krauss, \plb{190}{1987}{199}.

\bibitem{beta} L. Baudis {\it{et al.}}, 
\plb{407}{1997}{219}.

\bibitem{cobe} George F. Smoot {\it et al.}, 
\apjl{396}{1992}{L1}; C. L. Bennett {\it et al.}, 
\apjl{464}{1996}{1}.

\bibitem{leptogenesis} M. Fukugita and 
T. Yanagida, \plb{174}{1986}{45}; 
W. Buchm\"uller and M. Pl\"umacher, 
\plb{389}{1996}{73}. In the context of inflation see 
G. Lazarides and Q. Shafi, \plb{258}{1991}{305}; 
G. Lazarides, C. Panagiotakopoulos and Q. Shafi, 
\plb{315}{1993}{325}, (E) \ibid{317}{1993}{661}. 

\bibitem{sarkar} U. Sarkar, \prd{59}{1999}{031301}. 

\bibitem{covi} L. Covi, E. Roulet and F. Vissani, 
\plb{384}{1996}{169}. 

\bibitem{pilaftsis} A. Pilaftsis, \prd{56}{1997}{5431}.

\bibitem{ibanez} L. E. Ib\'a\~nez and F. Quevedo, 
\plb{283}{1992}{261}.

\bibitem{deuterium} S. Burles and D. Tytler, 
\apj{499}{1998}{699}; \ibid{507}{1998}{732}.

\bibitem{giunti} C. Giunti, hep-ph/9802201.

\bibitem{chooz} M. Apollonio {\it{et al.}}, 
\plb{420}{1998}{397}.

\bibitem{smirnov} Alexei Yu. Smirnov, hep-ph/9611465 and 
references therein.

\bibitem{neu} G. Lazarides, Q. Shafi and N. D. Vlachos, 
\plb{427}{1998}{53}.

\end{thebibliography}
\end{document}